\documentclass[12pt,english]{article}

\usepackage{geometry}
\usepackage{physics}
\usepackage{natbib}
\usepackage{enumitem}

\usepackage[titletoc,title]{appendix}

\usepackage{setspace}
\raggedright

\usepackage{sectsty}
\allsectionsfont{\normalsize\raggedright\centering}

\usepackage[]{tocloft}
\addtocontents{toc}{\cftpagenumbersoff{section}}
\addtocontents{toc}{\cftpagenumbersoff{subsection}}

\makeatletter
\renewcommand{\@cftmaketoctitle}{}
\makeatother

\usepackage[]{footmisc}

\usepackage{authblk}

\usepackage{graphicx}
\usepackage{babel}

\usepackage{amsmath}
\usepackage{amsfonts}

\usepackage{framed}

\usepackage{xifthen}


\numberwithin{equation}{section}

\usepackage{txfonts}
\usepackage[T1]{fontenc}

\begin{document}

\title{{\huge \textbf{On the objectivity of measurement outcomes}}}
\author{{\large \textbf{Elias Okon}}}
\date{}

\maketitle

\thispagestyle{empty}

\begin{abstract}
Recent arguments, involving entangled systems shared by sets of Wigner’s friend arrangements, allegedly show that the assumption that the experiments performed by the friends yield definite outcomes, is incompatible with quantum predictions. From this, it is concluded that the results of measurements cannot be thought of as being actual or objective. Here, I show that these arguments depend upon a mistaken assumption, regarding the (``mixed'') correlations between the results of the friends and those of ``the Wigners'', which leads to invalid predictions. It is not, then, that the assumption of definite outcomes leads to trouble, but that the results derived with such an assumption are contrasted with faulty predictions. Next, I explore the more famous no-go theorem by Frauchiger and Renner, on which these recent arguments are motivated. I show that, although it is cast in a different form, it also rests on the mistaken assumption regarding ``mixed'' correlations---rendering it invalid. Throughout, I illustrate my claims with explicit calculations within pilot-wave theory.
\end{abstract}

\mbox{} \\

\tableofcontents

\mbox{} \\

\section{Introduction}
A number of recent arguments seek to challenge the objectivity of measurement outcomes. More concretely, such arguments, involving scenarios in which entangled systems are shared by sets of Wigner’s friend arrangements, try to show that the assumption that the experiments performed by the friends yield definite outcomes, leads to a contradiction. From this, it is concluded that the results of (at least some) measurements, cannot be thought of as being actual or objective.

These new arguments employ experimental settings that have been used to derive Bell-type results, such as Bell's own, GHZ's or Hardy's. However, they place, on each wing of the experiments, a Wigner's friend arrangement, where the friend first measures inside a sealed room, with Wigner standing outside, and Wigner measures afterwards. This procedure allows to employ, on a single run, measurement settings that, in a standard Bell-type settings, would correspond to different possibilities. For this reason, these arguments are touted as having the added advantage of being formulated in terms of actual results of measurements---and not in terms of counterfactuals or potentially unreliable theoretical elements, such as Bell's $\lambda$'s or PBR's ontic states.

One such argument is reported in a talk by Pusey \citep{Pusey}, where it is ultimately traced to an informal argument offered by Masanes (see also \cite{H1}; \cite{H2}; \cite{H3}; \cite{Bub}; \cite{Leifer}). The proposed construction involves a Bell-type setting, in which the two spin-$\frac{1}{2}$ particles of a singlet are sent to two spatially separated laboratories. Each laboratory is then stipulated to contain a Wigner's friend arrangement where, first the friend, and then Wigner, perform measurements. With such an experiment, and by assuming that all measurements yield definite outcomes, an inconsistency is claimed to arise. From that, the objectivity of measurement results is challenged.

Another such argument is offered in \cite{Zuk}, where an extended Wigner's friend scenario is used, this time inspired by the GHZ arrangement---a three spin-$\frac{1}{2}$ particles variant of Bell's. The conclusion of the analysis is that any attempt to introduce an actual outcome for the friends' measurements, leads to a logical contradiction.

The previous works are both motivated, at least in part, by the more famous no-go theorems in \cite{FR}, and \cite{Bru1}. Ironically, though, while the Pusey-Masanes argument is described as easier to understand, but containing all the essential features of \cite{FR}, Żukowski and Markiewicz present their work as ``nullifying'' the ``apparent paradoxes'' of \cite{Bru1} and \cite{FR}, and correcting their mistakes.

Frauchiger and Renner consider an extended Wigner's friend scenario modeled after Hardy's paradox---yet another variant of Bell's arrangement---and a conflict is argued to arise from the assumption that quantum mechanics can be consistently applied to complex, macroscopic systems. They take their result as proving that ``quantum theory cannot consistently describe the use of itself''. Building on \cite{Bru2}, the first to employ a Bell-Wigner mash-up, \cite{Bru1} presents a ``no-go theorem for observer-independent facts''. The result is interpreted as showing that there can be no theory in which the results of both Wigner and the friend can jointly be considered as objective properties of the world.

This work offers a detailed examination of these arguments. In particular, I show that the first two arguments crucially depend upon a mistaken assumption, regarding the ``mixed'' correlations between the results of the friends and those of ``the Wigners''. It is not, then, that the assumption of definite outcomes leads to trouble; instead, such an assumption, \emph{coupled with the unwarranted assumption about mixed correlations}, leads to inconsistencies. I further argue that the unwarranted assumption regarding mixed correlations can be traced to a lack of recognition i) that in these arguments, hidden variables, with their inevitable contextual and non-local nature, are being (implicitly) postulated, and ii) that, in spite of such features of hidden variables, signaling is fully avoided in the scenarios considered. Regarding the Frauchiger-Renner theorem, I show that, although it seems to be presented in a different language, at the end, it also crucially depends on the mistaken assumption regarding mixed correlations. As a result,  the theorem is invalid.

My manuscript is organized as follows. In section \ref{M}, I present the Pusey-Masanes and the Żukowski-Markiewicz arguments and, in section \ref{MCA}, I show that they depend on an unwarranted assumption regarding mixed correlations. After that, in section \ref{FR}, I explore the Frauchiger-Renner theorem, on which the previous arguments are motivated. I show that it depends on the same unwarranted assumption regarding mixed correlations. Along the way, I illustrate my claims with explicit calculations in the context of pilot-wave theory, the best-developed hidden-variable theory (in the appendix \ref{PWM}, I collect the details of such calculations). Finally, in section \ref{Co}, I state my conclusions.

\section{The Pusey-Masanes and Żukowski-Markiewicz arguments}
\label{M}

The Pusey-Masanes (PM) argument employs an Bell-type experiment that starts with the two spin-$\frac{1}{2}$ particles of a singlet state being sent to two spatially separated laboratories. There, observers $\mathcal{A}_1$ and $\mathcal{B}_1$ perform spin measurements along directions $a_1$ and $b_1$, respectively; we denote these measurement results, which take values $+1$ or $-1$, by $A_1$ and $B_1$. Next, observers $\mathcal{A}_2$ and $\mathcal{B}_2$, who are outside of the laboratories of $\mathcal{A}_1$ and $\mathcal{B}_1$, respectively, come and \emph{undo} these initial measurements. The idea behind this is that, from the point of view of, say, $\mathcal{A}_2$, all that happens when $\mathcal{A}_1$ measures, is some complicated unitary transformation. Therefore, $\mathcal{A}_2$ can, at least in principle, come and apply the inverse of such a unitary to $\mathcal{A}_1$'s laboratory to nullify the measurement. The last step is for $\mathcal{A}_2$ and $\mathcal{B}_2$ to perform spin measurements on the original particles of the singlet, along directions $a_2$ and $b_2$, respectively; we denote the corresponding results by $A_2$ and $B_2$ (a schematic representation of the proposed experiment is presented in Figure 1).
\begin{figure}
\centering
\includegraphics[height=10cm]{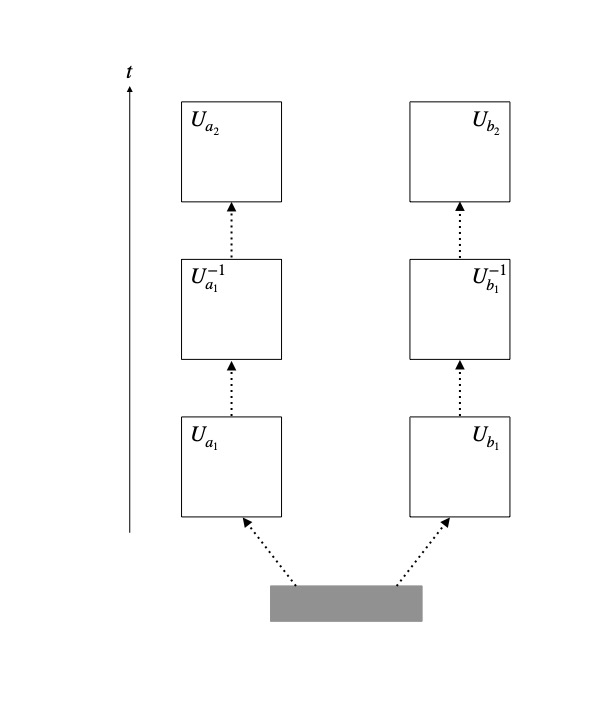} 
\caption{Schematic representation of the Pusey-Masanes experiment.}
\end{figure}

To develop the argument, we start by assuming that all the observers involved obtain objective results when they measure. That is, we assume that all the measurements performed yield definite, objective outcomes. As a result of such an assumption, there must exist a \emph{joint probability distribution} for all four results, $p(A_1,B_1,A_2,B_2)$. Given such a joint probability distribution, one can calculate the marginal probabilities $p_{\mathcal{A}_1\mathcal{B}_1}(A_1,B_1)$, $p_{\mathcal{B}_1\mathcal{A}_2}(B_1,A_2)$, $p_{\mathcal{A}_2\mathcal{B}_2}(A_2,B_2)$, $p_{\mathcal{A}_1\mathcal{B}_2}(A_1,B_2)$ and, with them, the expectation values of products of pairs of results $E_{\mathcal{A}_1\mathcal{B}_1}$, $E_{\mathcal{B}_1\mathcal{A}_2}$, $E_{\mathcal{A}_2\mathcal{B}_2}$, $E_{\mathcal{A}_1\mathcal{B}_2}$. Now, as shown in \cite{Fine}, the expectation values computed by these marginals necessarily satisfy the CHSH inequality
\begin{equation}
\label{chsh}
|E_{\mathcal{A}_1\mathcal{B}_1} + E_{\mathcal{B}_1\mathcal{A}_2} + E_{\mathcal{A}_2\mathcal{B}_2} - E_{\mathcal{A}_1\mathcal{B}_2}| \le 2 .
\end{equation}
That is, to assume a joint distribution for all four outcomes, leads to the same consequences than imposing Bell's locality. Of course, an important difference is that, in this case, all four measurements are actually performed in each run of the experiment.

Next, we employ quantum mechanics to make predictions for the expectation values in Eq. (\ref{chsh}). We start with $E^q_{\mathcal{A}_1\mathcal{B}_1}$, for which we notice that the situation exactly corresponds to a standard Bell scenario. In consequence,
\begin{equation}
E^q_{\mathcal{A}_1\mathcal{B}_1} = - \cos(a_1-b_1).
\end{equation}
Similarly, $E^q_{\mathcal{A}_2\mathcal{B}_2}$ seems easy to compute. The state starts being a singlet, then comes a big identity in its evolution, followed by standard Bell-type measurements. As a result,
\begin{equation}
\label{E2}
E^q_{\mathcal{A}_2\mathcal{B}_2} = - \cos(a_2-b_2).
\end{equation}

The ``mixed'' terms, with one measurement by $\mathcal{A}_1$ ($\mathcal{B}_1$) and the other by $\mathcal{B}_2$ ($\mathcal{A}_2$) seem more challenging. However, one might reason as follows. Say, $\mathcal{A}_1$, measures her particle. Then, to predict the result of $\mathcal{B}_2$, she updates the state according to the result obtained, and evolves the state on the other side. However, that evolution simply amounts to the identity, corresponding to the doing and undoing of the measurement. As a consequence, according to  $\mathcal{A}_1$, the relation of her particle to that of $\mathcal{B}_2$, is exactly the same than to the particle of $\mathcal{B}_1$, so the prediction for the mixed correlation must be the same as in the original Bell scenario.

Another way to put  this is that, for $\mathcal{A}_1$ to compute the correlation between her measurement and that of $\mathcal{B}_2$, she observes that, since the evolution from $\mathcal{B}_1$ to $\mathcal{B}_2$ is the identity, then the state of her particle and that of $\mathcal{B}_2$ is simply a singlet. Therefore, a straightforward application of Born's rule leads to the standard quantum correlations. From all this, it follows that 
\begin{equation}
\label{mix}
E^q_{\mathcal{A}_1\mathcal{B}_2} = - \cos(a_1-b_2), \qquad E^q_{\mathcal{B}_1\mathcal{A}_2} = - \cos(b_1-a_2).
\end{equation}

The problem, of course, is that we know that these quantum predictions can violate the CHSH inequality. That is, for adequate values of $a_1,b_1,a_2,b_2$,
\begin{equation}
\label{chshq}
|E^q_{\mathcal{A}_1\mathcal{B}_1} + E^q_{\mathcal{B}_1\mathcal{A}_2} + E^q_{\mathcal{A}_2\mathcal{B}_2} - E^q_{\mathcal{A}_1\mathcal{B}_2}| > 2 .
\end{equation}
We arrive, then, at a contradiction. In detail, the assumption that all measurements performed in the experiment yield definite, objective outcomes, which is what allowed for Eq. (\ref{chsh}) to be established, is found to be incompatible with the predictions of quantum mechanics in Eq. (\ref{chshq}).

From all this, it is concluded that quantum measurement outcomes cannot be taken to be objective, at least not in a straightforward way. That is, the argument is taken to entail that not every quantum measurement can be thought of as having a definite, objective, physical outcome. Moreover, the argument is read as implying that, at least in some sense, whatever maneuver is employed to deal with Bell's theorem, such a maneuver cannot be restricted to the microscopic level. That is, that the measure taken to address the CHSH violation in the Bell case, must also have an effect on actual, macroscopic experimental results.

The Żukowski-Markiewicz (ZM) argument \citep{Zuk}, is to the PM argument, what the GHZ scheme \citep{GHZ}, is to Bell's theorem. They consider a setting in which the three spin-$\frac{1}{2}$ particles of a GHZ state, given by
\begin{equation}
\label{GHZ}
|GHZ \rangle = \frac{1}{\sqrt{2}} \left[| +_z, +_z, +_z \rangle + | -_z, -_z, -_z \rangle \right] ,
\end{equation}
are distributed to three spatially separated laboratories, numbered $m=$1, 2, 3. In each of them, the particle is sent to a sealed room, where it is measured by the corresponding friend, $F_m$, in the $y$ basis. Such a measurement is assumed to involve a purely unitary evolution, in which the friend simply gets correlated with the measured particle. That is, in which $ | \text{ready} \rangle_{F_m} | \pm_y \rangle_m \rightarrow | \pm \rangle_{F_m} | \pm_y \rangle_m $. Still, the measurement is assumed to yield a definite outcome. After that, the corresponding ``Wigner'', $W_m$, measures the joint friend+particle system in the $x$ basis. This measurement is assumed to be irreversible.\footnote{\cite{Zuk} incorrectly states that \emph{decoherence} is able to explain such an irreversible process (see \cite{Less} for a detailed explanation of why this is not the case). In any case, such a misconception is immaterial for the description and evaluation of the argument.}

Before constructing the argument, we note that, unlike the PM experiment, this one does not involve the undoing of the friends' measurements. However, such a difference is not really substantial. This is because, in this case, ``the Wigners'' measure the composite system, friend+particle, and not only the particle, as in the PM case. This, together with the assumption that the friends' measurements only involve them getting correlated with the particles, implies that the procedures yield analogous results.

To develop the argument, the following fact is noted. Since the measurements of the friends are fully described by unitary transformations---in which the friends and the particles get entangled---then, the probabilities that ``the Wigners'' assign to their possible results are exactly the same than what they would assign to direct measurements of the particles, if the friends were fully removed from the arrangement. They take this to imply that the results of the friends' measurements must coincide with what ``the Wigners'' would have gotten, in case the friends were removed. This, in turn, is taken to imply, for example, that $P(r,s,t|n,n',n'')_{W_1,F_2,F_3} = P(r,s,t|n,n',n'')_{W_1,W_2,W_3}$ for  results $r,s,t$ (with values $\pm 1$) and settings $n, n', n''$. That is, that in the case in which all friends and all Wigners measure, the mixed probabilities for the possible results of some friends and some Wigners, and thus, the mixed correlations of those results, must mimic the probabilities and correlations of the results of all three Wigners. This, however, is shown to lead to trouble.

As is well know, the GHZ state implies a number of perfect correlations. In particular, the expectation value of the product of the results for the three particles satisfies
\begin{equation}
\label{GHZE}
E(x,y,y) = E(y,x,y) = E(y,y,x) = -E(x,x,x) = -1.
\end{equation}
So, if we denote the results of the Wigners and the friends by $w_m$ and $f_m$, we must have $w_1 f_2 f_3 = -1$, $f_1 w_2 f_3 = -1$,  $f_1 f_2 w_3 = -1$ and $w_1 w_2 w_3 = 1$. But this means that $(f_1 f_2 f_3)^2 = -1$, which is inconsistent with the fact that all results have values $\pm1$.

From this, it is concluded that the attempt to introduce actual outcomes for the measurements of the friends, leads to a logical contradiction. That is, that such measurements cannot be though of as being actual in any way.

\section{Mixed correlations}
\label{MCA}

What are we to make of these arguments? Do they really show that the objectivity of measurements must be called into question? Do we really have to throw away the sensible idea that well-conducted measurements yield definite, physical outcomes?

We start by noting that a key element of these arguments is the calculation of mixed correlations between results of the friends and those of ``the Wigners''. For instance, in the PM case, we had $E^q_{\mathcal{A}_1\mathcal{B}_2} = - \cos(a_1-b_2) $ and $E^q_{\mathcal{B}_1\mathcal{A}_2} = - \cos(b_1-a_2)$. Similarly, in the ZM case, it is argued that relations, such as  $P(r,s,t|n,n',n'')_{W_1,F_2,F_3} = P(r,s,t|n,n',n'')_{W_1,W_2,W_3}$, must obtain. And, of course, such mixed correlations were essential for the conclusion that the postulation of definite outcomes is incompatible with the quantum predictions for these experiments (i.e., the mixed correlations were essential in showing that such predictions violate CHSH, in the first case, and satisfy Eq. (\ref{GHZE}), even for mixed observers, in the second one). That is, the assumption that the mixed correlations---between results of the friends and those of ``the Wigners''---are equal to the standard quantum correlations among only the friends, or only ``the Wigners'', is required for these arguments to get off the ground.

Two key questions arise:
\begin{enumerate}
\item Are these mixed correlations the correct quantum predictions for these experiments?
\item Is there, or could there be, empirical evidence for such mixed correlations?
\end{enumerate}
Clearly, if the first question has a negative answer, then the alleged inconsistencies would never arise. Moreover, if the answer to the second question is also negative, then the consequences of the assumption of objective outcomes, independently of what the quantum predictions actually are, could not lead to empirical trouble.

To answer these questions we start by pointing out the obvious fact that, for these arguments to run, one has to assume that, at least during the measurements of the friends, the evolution of the quantum state is purely unitary. This is explicitly stated in the ZM scenario and is implicitly understood in the PM argument; otherwise, in the second case, the whole idea of undoing the measurements clearly does not go through. That is, it is only because it is assumed that the measurements performed by the first observers could be fully described by a complicated unitary, that it is possible to argue that such measurements could be undone by the application of another unitary---namely, the inverse of the one describing the measurements. In contrast, if one assumes that measurements involve some sort of breakdown of unitarity, the arguments simply do not work. From this, it clearly follows that theories that take this latter position, such as objective collapse models (\cite{GRW}; \cite{CSL}), are simply not at all affected by these arguments.

It is clear, then, that these arguments require the assumption that the evolution of the quantum state is always unitary. Does that mean that, at least, we can conclude that \emph{purely unitary} quantum theory is inconsistent with the assumption that measurements have definite outcomes? Not really. As we saw, the arguments presuppose 1) that measurements always yield definite results, and 2) that the evolution of the quantum state is always unitary. Now, in \cite{Tim} it is shown that those two assumptions are incompatible with another assumption, namely 3) that the physical description given by the quantum state is complete. In other words, the assumptions of these arguments necessarily imply that the quantum description is incomplete, so it must be supplemented in some way.

The point is that, a theory in which time evolution is always unitary, even during measurements, and in which the quantum state is assumed to be complete, is simply incompatible with the claim that measurements yield objective outcomes. That is, if the initial state of the system is a superposition of eigenstates of the observable to be measured, and if the interaction of the system with the measuring apparatus is purely unitary, then the final state of the composite system+apparatus will be a superposition of the possible results. If, moreover, the quantum state is assumed to be complete, then such a final superposition is simply incompatible with the claim that only one of the terms of the superposition can be taken as the actual, objective result of the measurement (for an explanation of why decoherence does not alter this conclusion, see \cite{Less}).

One might complain that the incompatibility result in \cite{Tim} depends upon assuming some level of objectivity or realism that goes against the spirit of the new type of arguments under discussion. That is, that the whole point of these new arguments is precisely to stay clear of theoretical constructs, such as hidden variables. However, if the objective of these new results is to \emph{argue against} an objective position, it would amount to begging the question to assume from the onset that such a position should not be taken.

The conclusion, then, is that the assumptions of the arguments imply that the quantum state must be supplemented by so-called hidden variables. Regarding these hidden variables, we know from the Kochen-Specker theorem \citep{KS}, that they must be \emph{contextual}, meaning that the results of at least some measurements must depend on the measurement context. Therefore, the presuppositions of these arguments establish that the outcome of measurements may depend, not only on the initial state of the system, but also on the particular way in which measurements are conducted. In particular, the outcome of an experiment may depend on which other observables are also being measured, even if remotely (of course, this possible non-local effect of remote joint measurements is a direct consequence of the unavoidable non-locality unearthed by Bell (\cite{Bell1964}; \cite{Bell1971}; \cite{Bell1976}; \cite{Bell1990}); in fact, a violation of locality, in the sense of Bell's theorem, already implies a violation of contextuality, in the sense of Kochen-Specker).

With these issues cleared up, we can go back to the two key questions regarding mixed correlations. We start with the first one: are such mixed correlations the correct quantum predictions? In order to answer, let us review what are the arguments behind the correlations offered. In the PM case, it was argued that, in order for, say, $\mathcal{A}_1$, to predict the result of $\mathcal{B}_2$, she updates the state according to her result and evolves the state on the other side. And, since such an evolution amounts to an identity, she makes the same predictions as in the original Bell scenario. That is, the idea is that, since the state of the particle when $\mathcal{B}_2$ measures is the same than when $\mathcal{B}_1$ measures, then $\mathcal{A}_1$'s correlation with both must also be the same. In the ZM argument, on the other hand, the mixed correlations are arrived at by employing counterfactual reasoning. First it is noted that the friends' results must coincide with what ``the Wigners'' would have gotten, in case the friends were removed. From that counterfactual, it is argued that, in the case in which all friends and all Wigners do measure, the mixed correlations for the results of some friends and some Wigners must coincide with the correlations of the results of all three Wigners alone.

These arguments might sound reasonable, but the recognition that hidden variables, with their inevitable contextual and non-local nature, are being (implicitly) postulated, renders them invalid. As we saw, in the PM case, the argument depended on the claim that, since the state of the particle when $\mathcal{B}_2$ measures is the same than when $\mathcal{B}_1$ measures, then $\mathcal{A}_1$'s correlation with both must also be the same. However, as explained above, the contextual nature of the setting implies that the outcomes may depend, not only on the initial state of the system, but also on which other observables are also being measured. As a result, from the fact that the quantum state is the same when $\mathcal{B}_1$ and $\mathcal{B}_2$ measure, does not follow that the correlations of those results, with that of $\mathcal{A}_1$, must coincide; the context of each measurement must be taken into account.

Regarding the other argument, the counterfactual, namely, that the friends' results must coincide with what ``the Wigners'' would have gotten, in case the friends were removed, seems quite reasonable. The problem, is that from it, it does not follow that, when everybody measures, the mixed correlations obtain. The issue, once more, is that the contextuality of the hidden variables gets in the way: what other measurements are being performed elsewhere is fully relevant to compute probabilities and correlations between measurements. 

To see all this in a concrete example, we consider the PM arrangement, from the point of view of the best-developed hidden-variable framework, the de Broglie-Bohm pilot-wave theory (see appendix \ref{PWM} for all the details). Such a framework proposes that a complete characterization of a particle system is given by its wave function, \emph{together} with the actual positions of the particles, which are taken to always possess well-defined values.

Now, according to the pilot-wave perspective, a particle can be assigned an objective, definite value of spin, in relation to the path through which it is deflected when traveling through a Stern-Gerlach device \citep{Nor2}. Moreover, such a value of spin can be conferred, independently of whether the particle is measured to be there. That is, the particle can be assigned a spin value, even if there is no process to correlate its position, after it passes the apparatus, with some sort of macroscopic measuring device. Finally, after sending a particle through a Stern-Gerlach apparatus, one could use reverse magnets to deflect the two separating components and recombine them to reconstitute the original beam. Even in that case, according to the pilot-wave perspective, a definite result can be associated with the measurement.

As a result of all this, from the pilot-wave perspective, the PM experiment amounts to sending the two particles of a singlet in opposite directions, measuring them along directions $a_1$ and $b_1$, respectively, recombining the beams of each particle, and measuring them along directions $a_2$ and $b_2$, respectively. It must be mentioned that, within pilot-wave, the order of operations, even if remote, is relevant for the calculation of the results. Therefore, in order to make concrete predictions, we must specify a particular order in which the operations are performed. Given the particular order described in the appendix (see Figure 3), the expectation values of pairs of results can be calculated to be
\begin{eqnarray}
\label{Es}
E^{p}_{\mathcal{A}_1\mathcal{B}_1} &=&  -\cos(a_1-b_1) \nonumber \\
E^{p}_{\mathcal{A}_2\mathcal{B}_2} &=&  -\cos(a_2-b_2) \nonumber \\
E^{p}_{\mathcal{A}_1\mathcal{B}_2} &=&  0 \nonumber \\
E^{p}_{\mathcal{B}_1\mathcal{A}_2} &=& -\left(1- \frac{2 \alpha}{\pi} \right)\cos(a_1-b_1) - \left(1-\frac{2 \beta}{\pi} \right) \cos(a_2-b_2)  \nonumber \\
& & - \frac{2}{\pi} \left\lbrace \alpha \cos(a_1-b_1) |\cos(a_2-b_2)| + \beta |\cos(a_1-b_1)| \cos(a_2-b_2) \right\rbrace \nonumber \\
& & + 2 \cos(a_1-b_1) \cos(a_2-b_2) \theta(\cos(a_1-b_1), \cos(a_2-b_2)) \nonumber \\
& & - 2 \cos(a_1-b_1) \cos(a_2-b_2) \theta(-\cos(a_1-b_1), -\cos(a_2-b_2)),
\end{eqnarray}
with $\alpha $ the angle between $a_1$ and $a_2$, $\beta $ the angle between $b_1$ and $b_2$ and  $\theta (x,y)$ the two-dimensional Heaviside step function, which is 1 only when both $x$ and $y$ are positive.

Given these results, several comments are in order. First, we note that, while the correlations between the first and second sets of measurements coincide with the standard quantum predictions, the mixed correlations do not satisfy Eq. (\ref{mix}). In fact, we know from Fine's theorem that the predictions of any framework that assigns definite values to all measurements, such as pilot-wave, must satisfy CHSH and, indeed, it can be checked that these results do so. We also note, from $E^{p}_{\mathcal{B}_1\mathcal{A}_2}$, the explicit non-locality and contextuality of the theory, as the correlation between the measurements of $\mathcal{B}_1$ and $\mathcal{A}_2$ depends on the angles employed by $\mathcal{A}_1$ and $\mathcal{B}_2$. Finally, we note that, even though these results would change if the order of operations changes, the new results would share with these results the relevant features mentioned above.

One might worry that the non-local character of $E^{p}_{\mathcal{B}_1\mathcal{A}_2}$ could offer the possibility of signaling. For instance, the correlations between the first and second measurements on each side are given by
\begin{eqnarray}
E^{p}_{\mathcal{A}_1\mathcal{A}_2} &=& \left(1-\frac{2 \alpha}{\pi} \right) \left( 1-|\cos(a_1-b_1)| \right) \nonumber \\
E^{p}_{\mathcal{B}_1\mathcal{B}_2} &=&  \left(1-\frac{2 \beta}{\pi} \right)  \left( 1-|\cos(a_2-b_2)| \right) .
\end{eqnarray}
Therefore, the correlation on each side gets modulated by measurements on the other. In particular, the correlation between, say, $\mathcal{A}_1$ and $\mathcal{A}_2$, can be made to appear or disappear, by adequate choices of $b_1$. It seems, then, that a protocol could be constructed in order for the $\mathcal{B}$-side to send a superluminal message to the $\mathcal{A}$-side. We must note, however, that, by construction, it is simply impossible to actually compare first and second measurements on a given side. The point is that, by stipulation, after the first measurements, the beams are recombined, which means that absolutely all records of such measurements are fully erased. We conclude that, after all, this scheme cannot be employed for superluminal signaling.

Going back to Eq. (\ref{Es}), it is clear that the pilot-wave predictions do not coincide with Eq. (\ref{mix}). Still, one could ask what are the ``correct'' standard, quantum predictions for the PM experiment. The truth is that \emph{standard quantum mechanics simply cannot make predictions for the mixed correlations in these experiments}. The point is that, as we saw above, these arguments presuppose, both, a fully unitary evolution during measurements (at least those of the friends), and the existence of definite results for all measurements. The problem is that, within the standard framework, for definite results to obtain, a breakdown of unitarity is required. Otherwise, one ends up with a superposition of possible outcomes and not a definite result. In other words, the presuppositions of the arguments are incompatible from the get-go with the standard framework and, as a result, it is simply meaningless to ask for the correct standard quantum predictions for these scenarios.

At this point, it seems useful to speculate on what led the proponents of these arguments to embrace the mistaken assumptions. That is, to speculate on what views on quantum theory, made them blind to the fact that the mixed correlations are, in fact, incorrect. I advance that the root of the problem can be found in three interrelated issues.

First, the fact that the proponents of these arguments, implicitly or explicitly, espouse an \emph{operational} reading of quantum theory, which makes attributing objectivity to measurement outcomes foreign, even meaningless. However, as mentioned above, if the whole point of these arguments is to argue against an objective stance, adopting a position which, effectively, does not allow for objective outcomes to obtain, amounts to begging the question.

The second issue is the fact that the operational position adopted, effectively ignores the so-called measurement problem, i.e., the inherent ambiguity of the standard framework regarding what happens during measurements. However, simply ignoring such a problem, does not make it go away. The point is that, it is this ambiguity of the standard framework, when applied to these complicated Wigner’s friends scenarios, which leads to seemingly inconsistent results. Then, by ignoring the measurement problem, the inconsistencies encountered are blamed on \emph{objectivity}, and not on the inconsistencies built into the perspective from the beginning. As explained above, standard quantum mechanics simply cannot make predictions for the mixed correlations in these experiments, and this is true even if one adopts an operational position.

Finally, even accepting the measurement problem, and conceding that hidden variables are presupposed, it could be argued that one could simply ignore them, and use the Born rule to make predictions. This, however, would be mistaken. The point is that, if one assumes i) purely unitary evolution during measurements and ii) that all measurements yield objective outcomes, then there are legitimate and illegitimate uses of the Born rule. The way to distinguish between them, is by employing a theory which is, in fact, consistent with those assumptions, i.e., a hidden-variable theory, such as pilot-wave.

Now, since for situations in which the standard framework is able to make concrete predictions, e.g., situations which do not involve Wigner’s friend scenarios, pilot-wave yields the same predictions as the standard framework, in those situations, the predictions of pilot-wave, coincide with the Born rule. There are, however, other scenarios, such as those leading to the mixed correlations in the PM setting, in which the standard framework is unable to make predictions; in those cases, the use of the Born rule would be illegitimate. As we saw above, Pilot-wave can make those predictions, but they do not have to conform to the Born rule. In fact, in \cite{Berndl}, using an extended Wigner's friend scenario modeled after Hardy's paradox, it is shown that, in pilot-wave (and related theories), the Born rule cannot be satisfied on all space-like hyperplanes.

Before moving on, it is instructive to consider the predictions, for the PM setting, of a an objective collapse model, in which all measurements are associated with an objective, irreversible collapse of the quantum state. In that case, the expectation values would be given by
\begin{eqnarray}
E^{c}_{\mathcal{A}_1\mathcal{B}_1} &=&  -\cos(a_1-b_1) \nonumber \\
E^{c}_{\mathcal{A}_2\mathcal{B}_2} &=&  -\cos\alpha \cos\beta\cos(a_1-b_1) \nonumber \\
E^{c}_{\mathcal{A}_1\mathcal{B}_2} &=&  - \cos\beta\cos(a_1-b_1) \nonumber \\
E^{c}_{\mathcal{B}_1\mathcal{A}_2} &=&  -\cos\alpha \cos(a_1-b_1) .
\end{eqnarray}
As expected, these values deviate from those in Eqs. (\ref{E2}) and (\ref{mix}), so the PM argument does not apply. These values also deviate from those in Eq. (\ref{Es}), so the PM experiment would discriminate between pilot-wave theory and objective collapse models (although simpler experiments could accomplish the same thing, e.g., \cite{Bassi}, \cite{Das}).

The second key question regarding the mixed correlations is whether there is, or there could be, empirical evidence for them. The issue is that, even if frameworks that assume objective outcomes for all measurements are theoretically safe, regarding not being incompatible with standard quantum predictions, they could be in trouble if the mixed correlations where empirically required. It is easy to see, though, that the mixed correlations cannot be empirically accessed. As we saw above, the undoing of the measurements implies erasure of all records so, by the time the second measurements are performed, there is nothing to compare them with. More generally, considering all possible time orders of the different measurements of, say, the PM arrangement, it is possible to see that, at least one expectation value cannot be measured (see \cite{Leifer} for an equivalent assertion). As a result, the corresponding CHSH inequality can never be shown to be violated empirically, so not satisfying the mixed correlations does not imply a clash with experiments. 

To see that, for any time order of events, at least one expectation value cannot be measured, we denote by $ e_1 \prec e_2$ when $e_2$ is on, or inside, the future light cone at $e_1$. Now, for the correlations between $\mathcal{A}_1$ and $\mathcal{B}_2$ to be empirically accessible, we need the undoing of the measurement on the $\mathcal{A}$ side to be to the future of the measurement of $\mathcal{B}_2$, i.e., we need $e_{b_2} \prec e_{a_1^{-1}}$. Otherwise, $\mathcal{B}_2$ would not have access to $\mathcal{A}_1$'s result. However, by construction of the experiment, it must be the case that $e_{a_1^{-1}} \prec e_{a_2} $ and $ e_{b_1^{-1}} \prec e_{b_2}$. That is, the second measurement on each side must be preceded by the undoing of the first measurement on that same side. The problem is that these three time orders imply that $e_{b_1^{-1}} \prec e_{a_2}$, which means that the correlations between $\mathcal{B}_1$ and $\mathcal{A}_2$ cannot be measured. We conclude that, if one of the mixed correlations can be measured, the other one cannot.

It is important to point out that, for those correlations that can be empirically accessed, hidden-variable theories have enough means, through contextuality and non-locality, to ensure that the predictions in such cases are in accordance with the standard framework (which, in such cases, is able to make predictions). This, indeed, can be shown to be the case for pilot-wave, the prime example of a hidden-variable theory.

In sum, not satisfying the mixed correlations does not lead to any sort of trouble: it neither implies a clash with standard predictions, nor it could lead to a conflict with experiments. Therefore, it is not that the assumption of definite outcomes leads to trouble, but that such an assumption, \emph{coupled with the unwarranted assumption regarding mixed correlations}, leads to inconsistent results. We conclude that these arguments fully fail in placing restrictions to theories that stipulate all measurements to lead to objective outcomes.

\section{The Frauchiger-Renner no-go theorem}
\label{FR}

In this section, we explore the no-go theorem in \cite{FR}, on which the previous arguments are motivated, and about which a lot has been written (\cite{baumann}; \cite{sudbery1}; \cite{sudbery2}; \cite{Drezet}; \cite{lazarovici}; \cite{tausk}; \cite{WCF}). In their work, Frauchiger and Renner present a result, which they take as proving that ``quantum theory cannot consistently describe the use of itself''. I confess that I do not quite understand what they mean by this phrase; in any case, what they actually do is to consider an extended Wigner's friend scenario, this time based o Hardy's variant of Bell's theorem (\cite{Hardy1}; \cite{Hardy2}), and the conflict is argued to arise from the assumption that quantum mechanics can be consistently applied to complex, macroscopic systems.

In more detail, the result is cast in the form of a no-go theorem, arguing for the mutual incompatibility of three assumptions: C, demanding consistency between different observers, S, establishing that measurements have single outcomes and Q, capturing the universal validity of quantum mechanics.\footnote{In more detail, the assumptions employed are as follows. C: if observer $O$ concludes that $O'$ is certain about the value of a property, then $O$ must also be certain about such a value; S: if $O$ is certain that the value of a property is $\xi$, then she must deny that she is certain that the value of that property is not $\xi$; Q: if $O$ assigns a state to a system to be measured, and that state assigns a Born probability of 1 to the result $\xi$, then $O$ must be certain that, after the measurement, the value of the measured property is $\xi$.} As I explained above, the Frauchiger-Renner (FR) experimental arrangement is, once more, an extended Bell-type scenario, this time modeled after Hardy's setting. There are, however, important differences between this argument and the previous ones, so it is useful to describe it in detail.

The arrangement contains four different agents, F, $\overline{\text{F}}$, W and $\overline{\text{W}}$, and two labs, L and $\overline{\text{L}}$. F is inside L and $\overline{\text{F}}$ inside $\overline{\text{L}}$ and W and $\overline{\text{W}}$ are outside of their respective labs and can perform measurements on them. Finally, there is a communication channel from $\overline{\text{L}}$ to L. The experiment runs in steps as follows (in \cite{FR}, these steps are repeated in rounds, but that is an unnecessary complication):
\begin{description}[font=\normalfont,labelindent=.5cm]
\item[\textbf{Step 1}:] $\overline{\text{F}}$ prepares a quantum coin in the state $\sqrt{\frac{1}{3}}\ket{h}+ \sqrt{\frac{2}{3}}\ket{t}$ and measures it. If she finds \emph{h}, she prepares an electron in the state $\ket{\downarrow}$, if she finds \emph{t}, she prepares it in the state $\ket{\rightarrow}= \frac{1}{\sqrt{2}} \left( \ket{\uparrow}+\ket{\downarrow} \right)$. She then sends the prepared electron to F.
\item[\textbf{Step 2}:] F measures the electron in the $\{\ket{\uparrow},\ket{\downarrow}\}$ basis.
\item[\textbf{Step 3}:] $\overline{\text{W}}$ measures $\overline{\text{L}}$ in the basis
$
\{\ket{o}_{\overline{\text{L}}}=\frac{1}{\sqrt{2}}\left[\ket{h}_{\overline{\text{L}}}-\ket{t}_{\overline{\text{L}}}\right],
\ket{f}_{\overline{\text{L}}}=\frac{1}{\sqrt{2}}\left[\ket{h}_{\overline{\text{L}}}+\ket{t}_{\overline{\text{L}}}\right] \}
$
(with $\ket{h}_{\overline{\text{L}}}$ and $\ket{t}_{\overline{\text{L}}}$ the states of $\overline{\text{L}}$ after $\overline{\text{F}}$ measures the coin and finds the corresponding result). Then she announces her result.
\item[\textbf{Step 4}:] W measures L in the basis
$
\{ \ket{o}_{\text{L}}=\frac{1}{\sqrt{2}}\left[\ket{\downarrow}_{\text{L}}-\ket{\uparrow}_{\text{L}}\right],
\ket{f}_{\text{L}}=\frac{1}{\sqrt{2}}\left[\ket{\downarrow}_{\text{L}}+\ket{\uparrow}_{\text{L}}\right] \}
$
(with $\ket{\downarrow}_{\text{L}}$ and $\ket{\uparrow}_{\text{L}}$ the states of L after F measures the spin and finds the corresponding result).
\end{description}

The experiment is then analyzed assuming C, S and Q. By doing so, several implications are considered (see Table 3 in \cite{FR}):
\begin{enumerate}
\item If $\overline{\text{F}}$ finds \emph{t}, then she sends F the electron in state $\ket{\rightarrow}$. If so, $\overline{\text{F}}$ reasons, after F measures, for W the state of L will be $\ket{f}_{\overline{\text{L}}}$, so W's measurement will result for sure in \emph{f}. That is:
\begin{equation}\label{I1}
\text{If $\overline{\text{F}}$ finds \emph{t}, she knows that W will find \emph{f} with certainty.}
\end{equation}
\item If F finds $\uparrow$, then she reasons that $\overline{\text{F}}$ must have gotten \emph{t}. But, from (\ref{I1}), this implies that W will find \emph{f} with certainty. Then, using C:
\begin{equation}\label{I2}
\text{If F finds $\uparrow$, then she knows that W will find \emph{f} with certainty.}
\end{equation}
\item According to $\overline{\text{W}}$, after step 2, the system $\overline{\text{L}}$+L is in the state
$
\sqrt{\frac{1}{3}}\ket{h}_{\overline{\text{L}}}\ket{\downarrow}_{\text{L}}+ \sqrt{\frac{2}{3}}\ket{t}_{\overline{\text{L}}}\ket{\rightarrow}_{\text{L}} .
$
But that state is orthogonal to $\ket{o}_{\overline{\text{L}}}\ket{\downarrow}_{\text{L}}$. As a result:
\begin{equation}\label{Ix}
\text{If $\overline{\text{W}}$ finds \emph{o}, then she knows with certainty that F found $\uparrow$.}
\end{equation}
So, employing (\ref{I2}), (\ref{Ix}) and C:
\begin{equation}\label{I3}
\text{If $\overline{\text{W}}$ finds \emph{o}, then she knows that W will find \emph{f} with certainty.}
\end{equation}
\item Finally, using (\ref{I3}) and C:
\begin{equation}\label{I4}
\text{If $\overline{\text{W}}$ announces she got \emph{o}, then W knows she will find \emph{f} with certainty.}
\end{equation}
\end{enumerate}

The problem is that, at the same time, according to W, the state of L+$\overline{\text{L}}$, after step 2, is given by
$
\sqrt{\frac{1}{3}}\ket{h}_{\overline{\text{L}}}\ket{\downarrow}_{\text{L}}+ \sqrt{\frac{2}{3}}\ket{t}_{\overline{\text{L}}}\ket{\rightarrow}_{\text{L}} ,
$
which can also be written as
\begin{equation}
 \frac{1}{2\sqrt{3}} \left[ \ket{o}_{\overline{\text{L}}}\ket{o}_{\text{L}} - \ket{o}_{\overline{\text{L}}} \ket{f}_{\text{L}}+\ket{f}_{\overline{\text{L}}} \ket{o}_{\text{L}} \right] + \frac{\sqrt{3}}{2}\ket{f}_{\overline{\text{L}}} \ket{f}_{\text{L}}.
\end{equation}
Given that $\ket{o}_{\overline{\text{L}}}\ket{o}_{\text{L}}$ has a non-zero coefficient, W concludes that it is possible for her to get the result \emph{o}, even if $\overline{\text{W}}$ also gets \emph{o}. But this is inconsistent with (\ref{I4}).

From all this, it is concluded that the conjunction of S, C and Q is inconsistent, which means that (at least) one of them, must be given up.

What to make of this argument? We start by noting that it is not presented in the same form as the previous ones. To begin with, it involves one of the friends using the result of a measurement to decide which state to send to the other. This complicated procedure, however, is easily seen to be equivalent to the friends sharing, as in the PM and ZM arguments, an entangled system. In this case, an entangled coin and electron, in the state
\begin{equation}
\frac{1}{\sqrt{3}} \left[ \ket{h}\ket{\downarrow} + \ket{t}\ket{\uparrow} + \ket{t}\ket{\downarrow} \right] ,
\end{equation}
which can be recognized as a particular instance of a Hardy pair (\cite{Hardy1}; \cite{Hardy2}). After that, both friends measure in the $\{h,t\}$  and $\{\uparrow,\downarrow\}$ bases, respectively and, finally, ``the Wigners'' measure in their respective $\{o,f\}$ bases. Put like this, the FR set-up can be seen to be the Hardy version of the PM and ZM set-ups. Note that, as in the ZM case, and unlike in the PM one, in this case ``the Wigners'' measure their whole labs, containing the system and the friend. However, we already saw that this is equivalent to the undoing of the measurements, followed by measurements of only the particles, considered in the PM argument.

Another important difference between the FR theorem and the PM and ZM arguments, is that the former is cast in terms of\emph{epistemic} elements, such agents expressing certainty on certain beliefs, agents announcing their results or agents changing their beliefs on hearing the results of others. The problem is that, such a description at an epistemic level, is not accompanied by a clear and explicit description of how, physically speaking, measurements are to be dealt with. For instance, in the PM and ZM arguments, it was made clear that the measurements of the friends were to involve purely unitary evolution, and to lead to well-defined results. In this case, in contrast, we are not directly informed about any of these issues. And, of course, the inherently vague nature of the the standard framework---its ambiguity as to what happens during \emph{measurements}---allows for a range of interpretations in these non-standard scenarios, so a detailed prescription is required. As a result, in order to dissect the FR argument, the rules actually employed are to be extracted from the reasoning presented.

Regarding the evolution during measurements, we note, for instance, that in order to derive (\ref{I1}), $\overline{\text{F}}$ must assume that W describes F's measurement in a purely unitary way. Similarly, to derive (\ref{Ix}), $ \overline{\text{W}}$ must describe the measurements of F and $\overline{\text{F}}$ in a unitary fashion. We conclude that, as in the PM and ZM arguments, the FR theorem (implicitly) assumes measurements to involve purely unitary evolution.

Regarding measurement results, even if it is not stated explicitly in \cite{FR}, all measurements are assumed to yield definite results. For instance, as we just saw, to derive (\ref{I1}), W describes F's measurement in a purely unitary way. However, in the next step, in order to derive (\ref{I2}), it is assumed that F's measurement yields a definite outcome.\footnote{The conjunction of S and C could be thought to lead to the same conclusion. However, the fact that they are written in a purely epistemic language (see the previous footnote), makes the connection insecure.} In sum, as in the PM and ZM arguments, the derivation of the FR theorem assumes, both, measurements to involve purely unitary evolution, and measurements to lead to definite outcomes.

After these observations, we are in position to evaluate the FR argument. The first comment is that, just as the arguments in section \ref{M}, the FR argument employs a Bell-type scenario---in this case Hardy's---and extends it with Wigner's friend arrangements to allow all possible measurement settings to be employed on a single run. The second comment is that, just as the arguments in section \ref{M}, the FR argument employs mixed correlations between the measurements of the Fs and the Ws, such as (\ref{I1}) and (\ref{Ix}), in order to derive a contradiction.

Now, are these mixed correlations legitimate? Of course not. For instance, to derive (\ref{I1}), it is assumed that F's measurement does not spoil the correlation between the results of $\overline{\text{F}}$ and W. Similarly, (\ref{Ix}) depends on assuming that $\overline{\text{F}}$'s measurement does not break the correlation between the results of F and  $\overline{\text{W}}$. However,  as we saw in detail in section \ref{MCA}, this is incompatible with the (implicit) assumptions behind the theorem. As a result, these correlations are simply invalid. The upshot, of course, is that the whole FR theorem is unsound.

As in the PM case, we can illustrate all this with explicit calculations within pilot-wave theory (see appendix \ref{PWM} for details). In particular, we can compute the full joint probability for all four results, which includes the fact that
\begin{equation}
P_{\overline{\text{F}}FW\overline{\text{W}}}(t,\downarrow,o,o)=\frac{1}{9} .
\end{equation}
But this clearly shows that, if one takes seriously the (implicit) assumptions behind the theorem, then both (\ref{I1}) and (\ref{Ix}) are simply not satisfied. This is because the fact that this probability is different from zero, directly implies that it is neither the case that $\overline{\text{F}}$ finding \emph{t} implies W finding \emph{f}, nor that $\overline{\text{W}}$ finding \emph{o} implies F finding $\uparrow$. And, of course, without these correlations, no contradiction arises.

As mentioned above, a lot has been written about the FR theorem. In particular, in works such as \cite{baumann}, \cite{sudbery1}, \cite{sudbery2}, \cite{Drezet}, \cite{lazarovici}, \cite{tausk} and \cite{WCF}, it has been pointed out that the FR argument depends upon a hidden assumption. There is, however, no consensus regarding the exact nature of such an assumption. For instance, it has been argued that the theorem rests on an assumption to the effect that, when a measurement is carried out inside of a closed lab, such a measurement leads to a collapse for inside observers, but it does not lead to a collapse for outside observers. While I take such observations to be on the right track, I believe that the characterization of the hidden assumption offered in this work, in terms of mixed correlations, manages to fully uncover the true nature of the required additional assumption.

Moreover, and more importantly, this work is the first to show that this hidden assumption, is not only needed for the arguments to run, but that it is, in fact, invalid. This, of course, makes for a crucial difference: \emph{the FR argument is not only found to be limited, it is found to be false}. Finally, by exploring the FR argument in conjunction with similar arguments, the present work is not only able to shed light on the FR result, but to reveal the underlying structure of the family of arguments employing extended Wigner's friend scenarios to challenge the objectivity of quantum mechanics. In doing so, it shows why such results fully fail in their mission.

\section{Conclusions}
\label{Co}

Recent arguments, involving entangled systems shared by sets of Wigner’s friend arrangements, aim at calling into question the objectivity of measurement results. In particular, they purportedly show that the assumption that quantum measurements yield definite, objective outcomes, is incompatible with quantum predictions.

In this work, I show that such arguments depend upon a mistaken assumption---regarding the mixed correlations between the results of the friends and those of ``the Wigners''---which leads to invalid predictions. It is not, then, that the assumption of definite outcomes leads to trouble, but that the results derived with such an assumption are contrasted with faulty predictions. I also explore the more famous no-go theorem by Frauchiger and Renner, on which these recent arguments are motivated. I argue that it depends upon the same mistaken assumption about mixed correlations. I conclude that none of these results is able to impose interesting constraints on the assumption that measurements lead to objective outcomes.


\begin{appendices}
\section{Pilot-wave analysis of the Pusey-Masanes and Frauchiger-Renner experiments}
\label{PWM}
In this appendix, I offer a detailed analysis of the Pusey-Masanes (PM) and the Frauchiger-Renner (FR) experimental settings, in the context of the de Broglie-Bohm pilot-wave theory \citep{Bohm}. For this, I rely heavily on the pilot-wave treatment of spin presented in \cite{Nor2}.

Pilot-wave theory is the best-developed example of a hidden-variable model. The framework proposes that a complete characterization of a $N$-particle system is given by its wave function $ \Psi (x, t) $, \emph{together} with the actual positions of the particles \linebreak $\{\mathbf{X}_1 (t),\mathbf{X}_2 (t),\dots,\mathbf{X}_N (t)\}$, which are taken to always possess well-defined values. The wave function is postulated to satisfy at all times the usual Schr\"odinger equation
 \begin{equation}
 i \hbar \frac{\partial \Psi}{ \partial t } = \hat{H} \Psi ,
 \end{equation}
and the positions to evolve according to the deterministic ``guiding'' equation
\begin{equation}
\frac{d \mathbf{X}_k(t)}{dt} = \left. \frac{\sum_\nu \mathbf{j}_{k,\nu} (\mathbf{x}_1,\mathbf{x}_2,\dots,\mathbf{x}_N,t)}{\sum_\nu\rho_\nu(\mathbf{x}_1,\mathbf{x}_2,\dots,\mathbf{x}_N,t)}  \right\rvert_{\mathbf{x}=\mathbf{X}(t)},
\label{guide}
\end{equation}
with $\mathbf{j}_{k,\nu} = (\hbar/2mi) (\Psi_\nu^* \nabla_k \Psi_\nu - \Psi_\nu \nabla_k \Psi_\nu^*) $, $\rho_\nu = \Psi_\nu^* \Psi_\nu$ and $\nu$ a spin index, if available. Finally, the initial particle positions are assumed to be random, with probability distribution at $t=0$ given by
\begin{equation}
\label{SP}
P (\mathbf{X} = \mathbf{x}) = \sum_\nu | \Psi_\nu (\mathbf{x}) |^2.
\end{equation}

Before analyzing the full PM and FR settings, we consider a single spin-$\frac{1}{2}$ particle, traveling along the $y$ axis, which is sent through a Stern-Gerlach apparatus that produces a magnetic field $\mathbf{B} \approx b z \hat{z}$. Following \cite{Nor2}, we assume that the magnetic field is very strong, but is non-zero only in an extremely small region around $y=0$. This allows for the standard Stern-Gerlach interaction Hamiltonian to be well-approximated by $\hat{H} = - \mu b z \delta(y) \sigma_z$.

We assume the particle's initial wave function to be given by what \cite{Nor1} calls a ``plane-wave packet''. Such an incident wave packet is assumed to be nearly constant over a tubular region of length $L$ along the $y$ direction and radius $r$ centered at the origin of the $x-z$ plane, and to be zero outside such a region. Moreover, the particle is assumed to initially propagate along the $+y$-direction, with a reasonably sharply defined wave number $k$, such that $\lambda = 2 \pi/k \ll L$. This guarantees that, to a very good approximation, the packet will maintain its shape and travel at the appropriate group velocity.

As shown in \cite{Nor2}, by assuming that the impulse imparted to the particle by the apparatus, $\kappa$, is small compared to $k$, and that the width $2 r$ of the wave packet is small compared to $1/\kappa$, one can write
\begin{equation}
  \Psi(x,y,z) \propto
    \begin{cases}
       e^{i k y} (c_+ \chi_+ + c_- \chi_- ) & y<0 \\
       e^{i k' y} (e^{+i \kappa y} c_+ \chi_+ + e^{-i \kappa y} c_- \chi_- ) & y>0
    \end{cases}       
\end{equation}
with $k' \approx k$, $\chi_\pm$ eigenstates of $S_z$ and $c_\pm$ the weights of such eigenstates in the initial state. Of course, the above expression for $y<0$ is valid only where the wave function has support and, for $y>0$, in the region where the two separating components of the incident wave function overlap.

This wavefunction, together with the guiding equation, can be used to calculate the particle trajectories. For $y<0$ one obtains
\begin{equation}
\label{G1}
\frac{d \mathbf{X}(t)}{dt} = \frac{\hbar k}{m} \hat{y},
\end{equation}
which means that, no matter where inside the incident packet the particle is located, it will simply travel with the packet's group velocity. For $y>0$ one obtains
\begin{equation}
\label{G2}
\frac{d \mathbf{X}(t)}{dt} = \frac{\hbar k'}{m} \hat{y} + \frac{\hbar \kappa}{m} (|c_+|^2 +|c_-|^2) \hat{z},
\end{equation}
which amounts to a velocity in the $z$-direction equal to the weighted average of the group velocities of the two components of the wave function emerging out of the Stern-Gerlach apparatus.

Eq. (\ref{SP}), together with the fact that the initial wave function is assumed to be constant where it does not vanish, implies that the particle is equally likely to be located at any point inside the packet. Moreover, the pilot-wave dynamics is easily shown to preserve the initial distribution of Eq. (\ref{SP}), which means that the probability for the particle to end up in the upper (lower) path, and thus to be assigned spin-up (spin-down) when detected, is given by $c_+(c_-)$. To demonstrate this explicitly, one can analyze in detail what regions of the cross-section of the incoming packet result in the particle deflecting up or down; the result of such an analysis is given in Figure 2 (see \cite{Nor2} for details).
\begin{figure}
\centering
\includegraphics[height=8cm]{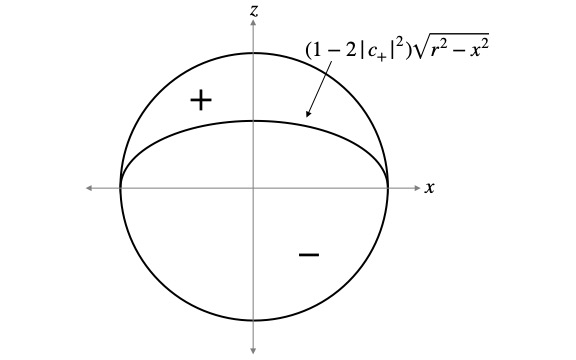} 
\caption{Regions of the cross-section of the incoming packet resulting in the particle deflecting up or down, upon encountering a Stern-Gerlach apparatus oriented along $z$.}
\end{figure}

At this point, it is important to point out that, according to the pilot-wave perspective, a particle could be assigned an objective, definite value of spin, according to the path through which it is deflected, independently of whether it is detected, observed or measured to be there. That is, the particle can be assigned a spin value, even if there is no process to correlate its position, after it passes the Stern-Gerlach device, with some sort of macroscopic measuring apparatus. Moreover, it is clear that, after sending a particle through a Stern-Gerlach device, one could use reverse magnets to deflect the two separating components and recombine them to reconstitute the original beam. Even in that case, according to the pilot-wave perspective, a definite result can be associated with the measurement. Finally, it is clear from Eqs. (\ref{G1}) and (\ref{G2}) that, to the level of the approximations employed, if the recombination is performed, after it, the $(x,z)$ position of the particle inside the wave packet will be exactly the position it had before it went through the apparatus.

The last element we need to review, before exploring the PM and FR settings, is the pilot-wave treatment of pairs of entangled particles. The first point to mention is that, as any other empirically viable hidden-variable theory, pilot-wave is non-local and contextual. In particular, such features imply that the results of an experiment may depend on what other measurements are being performed elsewhere. Moreover, the explicit non-locality of the framework requires an objective notion of simultaneity, which allows for the order of space-like separated events to be relevant for the calculation of the results of an experiment.

Consider the two spin-$\frac{1}{2}$ particles of a singlet, propagating from the origin along the $y$ axis, in opposite directions. Assume that the particles' initial wave functions are the plane wave packets described above. Suppose, moreover, two Stern-Gerlach apparatuses positioned on the $y$ axis, along the paths of the two particles. As explained in \cite{Nor2}, if particle 1 encounters its Stern-Gerlach apparatus first, then, whether it is deflected up or down (relative to the Stern-Gerlach orientation), is fully determined by the position of particle 1: if it is in the upper half of the wave packet, it will deflect up and, if it is in the lower half of the wave packet, it will deflect down. What happens when particle 2 is then measured? It can be shown that the behavior of particle 2, when measured, depends on where particle 1 ended up. In particular, if particle 1 is found with a certain spin, then particle 2 will effectively be guided by a wave function that corresponds to a wave function that ``collapsed'' according to the result obtained for particle 1. For instance, if both measurements are along the same axis, and if particle 1 is deflected up, then particle 2 will be deflected down, independently of its position---that is, even if its position is such that, if measured first, it would have deflected up. 

We are finally in position to analyze the experiments; we start with the PM case. We recall that it consists of the two spin-$\frac{1}{2}$ particles of a singlet being sent to two spatially separated laboratories, where observers $\mathcal{A}_1$ and $\mathcal{B}_1$ perform spin measurements along directions $a_1$ and $b_1$, respectively. Next, observers $\mathcal{A}_2$ and $\mathcal{B}_2$ \emph{undo} these initial measurements and perform measurements along directions $a_2$ and $b_2$, respectively. From the pilot-wave perspective, this experiment can be carried out without actually involving observers who perform and undo measurements. As I mentioned above, according to pilot-wave theory, when a particle goes through a Stern-Gerlach device, it can be though of as yielding a definite result for its spin, independently of whether the result is amplified or observed. Moreover, by recombining the beams, one achieves what is envisioned when the PM argument talks about the measurements being undone.

Therefore, from the pilot-wave perspective, the PM experiment amounts to sending the two particles of a singlet, $a$ and $b$, in opposite directions, measuring them along directions $a_1$ and $b_1$, respectively, recombining the beams, and measuring them along directions $a_2$ and $b_2$, respectively. We mentioned above, though, that the order of operations, even if remote, is relevant for the calculation of results. Therefore, in order to make concrete predictions, we must specify a particular order in which the operations are performed. For concreteness, we assume the following order: $t_1$) particle $a$ is measured along $a_1$; $t_2$) particle $b$ is measured along $b_1$; $t_3$) particle $b$ is recombined; $t_4$) particle $a$ is recombined; $t_5$) particle $b$ is measured along $b_2$ and $t_6$)  particle $a$ is measured along $a_2$ (see Figure 3).
\begin{figure}
\centering
\includegraphics[height=9cm]{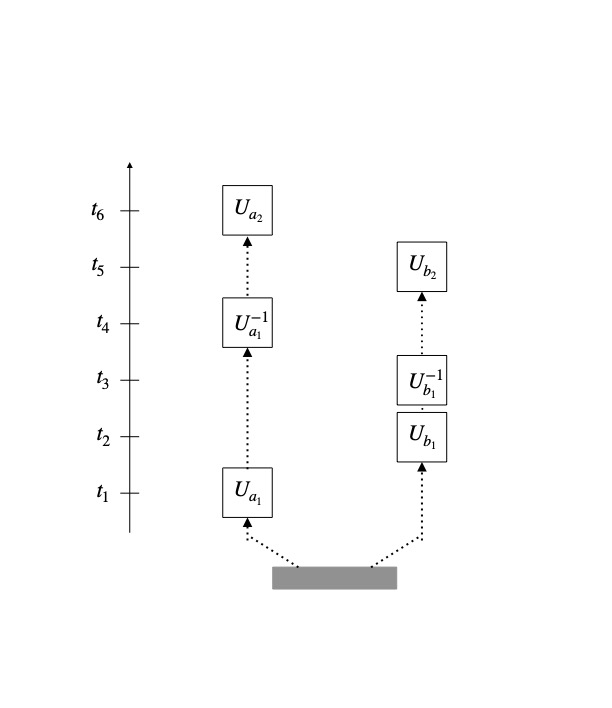} 
\caption{Schematic representation of the Pusey-Masanes experiment, with the particular order of operations used for the calculations.}
\end{figure}

Given what I have said so far, it can be shown that the possible results of the measurements depend on the initial positions of the particles according to Figure 4. Moreover, the sizes of the areas, corresponding to the different possible results, are given by
\begin{eqnarray}
a_{++} &=&  \frac{r^2}{2} \left[ (\pi-\alpha) + \alpha |2 \omega_a^\pm-1| - \pi \theta(1-2 \omega_a^\pm) (1-2\omega_a^\pm) \right] \nonumber \\
a_{+-} &=&  \frac{r^2}{2} \left[ \alpha - \alpha |2 \omega_a^\pm-1| + \pi \theta(1-2 \omega_a^\pm) (1-2\omega_a^\pm) \right] \nonumber \\
a_{-+} &=&  \frac{r^2}{2} \left[ \alpha - \alpha |2 \omega_a^\pm-1| + \pi \theta(2 \omega_a^\pm-1) (2\omega_a^\pm-1) \right] \nonumber \\
a_{--} &=&  \frac{r^2}{2} \left[ (\pi-\alpha) + \alpha |2 \omega_a^\pm-1| - \pi \theta(2 \omega_a^\pm-1) (2\omega_a^\pm-1) \right] \nonumber \\
b_{++} &=&  \frac{r^2}{2} \left[ (\pi-\beta) + \beta |2 \omega_b^\pm-1| - \pi \theta(1-2 \omega_b^\pm) (1-2\omega_b^\pm) \right] \nonumber \\
b_{+-} &=&  \frac{r^2}{2} \left[ \beta - \beta |2 \omega_b^\pm-1| + \pi \theta(2 \omega_b^\pm-1) (2\omega_b^\pm-1) \right] \nonumber \\
b_{-+} &=&  \frac{r^2}{2} \left[ \beta - \beta |2 \omega_b^\pm-1| + \pi \theta(1-2 \omega_b^\pm) (1-2\omega_b^\pm) \right] \nonumber \\
b_{--} &=&  \frac{r^2}{2} \left[ (\pi-\beta) + \beta |2 \omega_b^\pm-1| - \pi \theta(2 \omega_b^\pm-1) (2\omega_b^\pm-1) \right] , 
\end{eqnarray}
where $\theta (x,y)$ is the two-dimensional Heaviside step function, which is 1 only when both $x$ and $y$ are positive, $\alpha $ the angle between $a_1$ and $a_2$, $\beta $ the angle between $b_1$ and $b_2$, and with $\omega_a^\pm $ and $\omega_b^\pm $ determined by the results of the measurements along $b_2$ and $a_1$, respectively, according to
\begin{eqnarray}
\omega_a^+ &=& \sin^2(\frac{a_2-b_2}{2})  \nonumber \\
\omega_a^- &=&  \cos^2(\frac{a_2-b_2}{2}) \nonumber \\
\omega_b^+ &=& \sin^2(\frac{a_1-b_1}{2}) \nonumber \\
\omega_b^- &=&  \cos^2(\frac{a_1-b_1}{2}) .
\end{eqnarray}
\begin{figure}
\centering
\includegraphics[height=9cm]{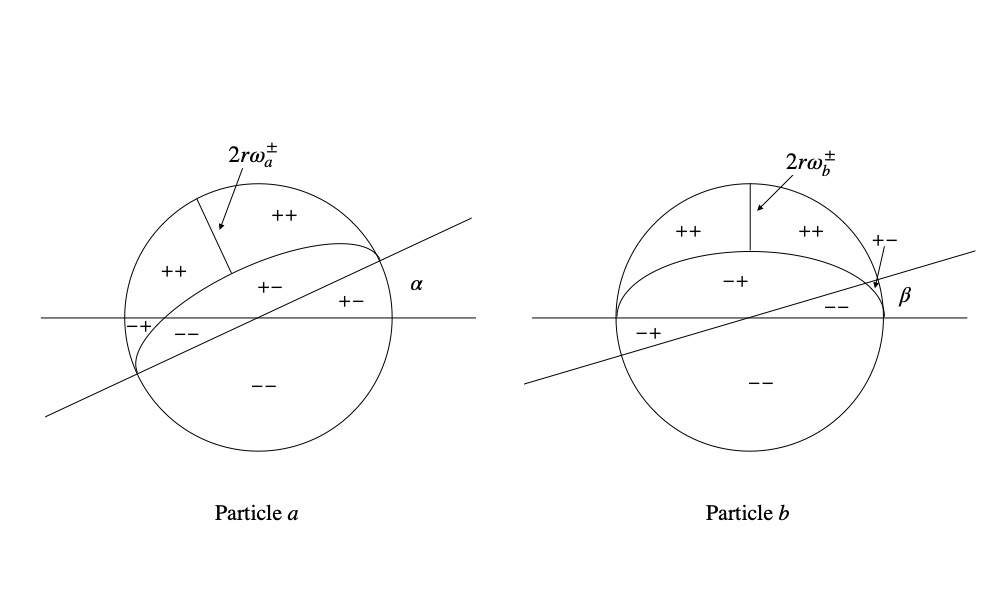} 
\caption{Regions of the cross-sections of the incoming packets, resulting in the particles deflecting up or down in their pair of measurements.}
\end{figure}

From all this, one can compute the joint distribution for all four measurements, then compute the marginals for pairs of results and, finally, the expectation values. The results are given by
\begin{eqnarray}
E^{p}_{\mathcal{A}_1\mathcal{B}_1} &=&  -\cos(a_1-b_1) \nonumber \\
E^{p}_{\mathcal{A}_2\mathcal{B}_2} &=&  -\cos(a_2-b_2) \nonumber \\
E^{p}_{\mathcal{A}_1\mathcal{B}_2} &=&  0 \nonumber \\
E^{p}_{\mathcal{B}_1\mathcal{A}_2} &=& -\left(1- \frac{2 \alpha}{\pi} \right)\cos(a_1-b_1) - \left(1-\frac{2 \beta}{\pi} \right) \cos(a_2-b_2)  \nonumber \\
& & - \frac{2}{\pi} \left\lbrace \alpha \cos(a_1-b_1) |\cos(a_2-b_2)| + \beta |\cos(a_1-b_1)| \cos(a_2-b_2) \right\rbrace \nonumber \\
& & + 2 \cos(a_1-b_1) \cos(a_2-b_2) \theta(\cos(a_1-b_1), \cos(a_2-b_2)) \nonumber \\
& & - 2 \cos(a_1-b_1) \cos(a_2-b_2) \theta(-\cos(a_1-b_1), -\cos(a_2-b_2))  ,
\end{eqnarray}
where the $p$ superscript indicates that these are the results, according to pilot-wave theory (given the particular order of operations chosen).

Regarding these results, we note that the first two correlation coefficients, exactly match the standard quantum prediction, which makes perfect sense. Such correlations can be calculated with the standard framework, and we know that, in such a case, the pilot-wave predictions reproduce the standard predictions. The third value, the one that is zero, also makes perfect sense. To begin with, the $a_1$ result is fully determined by the initial position of particle $a$. Moreover, by the time the $b_2$ measurement occurs, both the $a_1$ and $b_1$ measurements have been completely undone. Therefore, the $b_2$ measurement happens on a ``fresh'' singlet state, and its outcome is fully determined by the initial position of particle $b$. Finally, the initial positions of particles $a$ and $b$ are uncorrelated, from which it follows that the $a_1$ and $b_2$ results are uncorrelated. Regarding the fourth value, while it is relatively easy to see that such a correlation is neither zero, nor equal to the standard value, apart from the actual calculations, I do not have any insights regarding the actual value obtained.

In sum, we see that the expectation values for the first and second sets of measurements are, as expected, proportional to the cosine of the angle between the measurement axes. We see, however, that the mixed correlations depart from such a behaviour. This, of course, is a necessary consequence of Fine's theorem and, indeed, it can be checked that the first four values above, satisfy the CHSH inequality.

The correlations between the first and second measurements on each particle are given by
\begin{eqnarray}
E^{p}_{\mathcal{A}_1\mathcal{A}_2} &=& \left(1-\frac{2 \alpha}{\pi} \right) \left( 1-|\cos(b_1-a_1)| \right) \nonumber \\
E^{p}_{\mathcal{B}_1\mathcal{B}_2} &=&  \left(1-\frac{2 \beta}{\pi} \right)  \left( 1-|\cos(b_2-a_2)| \right) .
\end{eqnarray}
We see that these values explicitly show the non-locality and contextuality of the framework, in the sense that the correlation between the measurements on one side can be modulated by the selection of the orientation of a measurement on the other.

Next we analyze the FR arrangement. It contains four agents, F, $\overline{\text{F}}$, W and $\overline{\text{W}}$, and two labs, L and $\overline{\text{L}}$, with a communication channel between them. The experiment then proceeds as follows. First, Inside $\overline{\text{L}}$, $\overline{\text{F}}$ prepares a quantum coin in the state $\sqrt{\frac{1}{3}}\ket{h}+ \sqrt{\frac{2}{3}}\ket{t}$ and measures it. If she finds \emph{h}, she sends F an electron in the state $\ket{\downarrow}$, if she finds \emph{t}, she sends it in the state $\ket{\rightarrow}= \frac{1}{\sqrt{2}} \left[ \ket{\uparrow}+\ket{\downarrow} \right]$. Then, inside L, F measures the electron in the $\{\ket{\uparrow},\ket{\downarrow}\}$ basis. Next, $\overline{\text{W}}$ measures $\overline{\text{L}}$ in the basis
$
\{ \ket{o}_{\overline{\text{L}}}=\frac{1}{\sqrt{2}}\left[\ket{h}_{\overline{\text{L}}}-\ket{t}_{\overline{\text{L}}}\right], \ket{f}_{\overline{\text{L}}}=\frac{1}{\sqrt{2}}\left[\ket{h}_{\overline{\text{L}}}+\ket{t}_{\overline{\text{L}}}\right] \}
$
(with $\ket{h}_{\overline{\text{L}}}$ and $\ket{t}_{\overline{\text{L}}}$ the states of $\overline{\text{L}}$ after $\overline{\text{F}}$ measures the coin and finds the corresponding result), and announces her result. Finally, W measures L in the basis
$
\{ \ket{o}_{\text{L}}=\frac{1}{\sqrt{2}}\left[\ket{\downarrow}_{\text{L}}-\ket{\uparrow}_{\text{L}}\right], \ket{f}_{\text{L}}=\frac{1}{\sqrt{2}}\left[\ket{\downarrow}_{\text{L}}+\ket{\uparrow}_{\text{L}}\right] \}
$
(with $\ket{\downarrow}_{\text{L}}$ and $\ket{\uparrow}_{\text{L}}$ the states of L after F measures the spin and finds the corresponding result).

Now, as I explained in section \ref{FR}, the procedure in which $\overline{\text{F}}$ measures the coin, and sends the electron to F, is equivalent to $\overline{\text{F}}$ and F sharing an entangled coin-electron pair in the state
\begin{equation}
\label{ce}
\frac{1}{\sqrt{3}} \left[ \ket{h}\ket{\downarrow} + \ket{t}\ket{\uparrow} + \ket{t}\ket{\downarrow} \right] .
\end{equation}
Moreover, as I also explained above, given the assumption that measurements are to be described by a purely unitary evolution, the measurements of the Ws on their respective labs are equivalent to the initial measurements by the Fs being \emph{undone}, as in the PM case, with new measurements being performed directly on the coin and electron.

Given all this, the FR experiment is equivalent to the following procedure. $\overline{\text{F}}$ and F share an entangled coin-electron pair in the state (\ref{ce}). Then, $\overline{\text{F}}$ measures the coin in the $\{\ket{h},\ket{t}\}$ basis and F measures the electron in the $\{\ket{\uparrow},\ket{\downarrow}\}$ basis. Next, the coin and electron measurements are undone. Finally, they are measured again, the coin in the $\{\ket{f}_c=\frac{1}{\sqrt{2}}\left[\ket{h}+\ket{t}\right],\ket{o}_c=\frac{1}{\sqrt{2}}\left[\ket{h}-\ket{t}\right]\}$ basis, and the electron in the $\{\ket{f}_e=\frac{1}{\sqrt{2}}\left[\ket{\downarrow}+\ket{\uparrow}\right],\ket{o}_e=\frac{1}{\sqrt{2}}\left[\ket{\downarrow}-\ket{\uparrow}\right]\}$ basis.

Now, if the coin is substituted by an electron, given the techniques described above, this procedure can easily be modeled within pilot-wave theory. If one does so, and chooses the time order of operations analogous to the one used in the PM case, the full joint probability distribution over the results can be calculated. These probabilities are given by
\begin{eqnarray}
P_{\overline{\text{F}}FW\overline{\text{W}}}(h,\downarrow,f,o) &=& 1/9 \nonumber \\
P_{\overline{\text{F}}FW\overline{\text{W}}}(h,\downarrow,o,f) &=& 1/9 \nonumber \\
P_{\overline{\text{F}}FW\overline{\text{W}}}(h,\downarrow,o,o) &=& 1/9 \nonumber \\
P_{\overline{\text{F}}FW\overline{\text{W}}}(t,\uparrow,f,o) &=& 1/9 \nonumber \\
P_{\overline{\text{F}}FW\overline{\text{W}}}(t,\uparrow,o,f) &=& 1/9 \nonumber \\
P_{\overline{\text{F}}FW\overline{\text{W}}}(t,\uparrow,o,o) &=& 1/9 \nonumber \\
P_{\overline{\text{F}}FW\overline{\text{W}}}(t,\downarrow,f,o) &=& 1/9 \nonumber \\
P_{\overline{\text{F}}FW\overline{\text{W}}}(t,\downarrow,o,f) &=& 1/9 \nonumber \\
P_{\overline{\text{F}}FW\overline{\text{W}}}(t,\downarrow,o,o) &=& 1/9 ,
\end{eqnarray}
with all other terms equal to zero.

\end{appendices}
\section*{Acknowledgments}

I thank Ricardo Muciño and Travis Norsen for fruitful conversations, and two anonymous referees for valuable comments. I acknowledge support from UNAM-PAPIIT (grant IN102219) and CONACYT (grant 140630).


\begin{flushright}
\emph{
  Elias Okon\\
  Institute for Philosophical Research\\
  National Autonomous University of Mexico\\
  Circuito Maestro Mario de la Cueva s/n\\
  C.U., 04510, CDMX, México.\\
  eokon@filosoficas.unam.mx
}
\end{flushright}




\end{document}